\documentclass[twocolumn,pra,aps,superscriptaddress]{revtex4}

\usepackage{color}
\usepackage{epsfig}
\usepackage{latexsym}
\usepackage{amssymb}
\usepackage{amsmath}
\usepackage{algorithm}
\usepackage{algorithmic}
\usepackage{wrapfig}
\usepackage[apple mac]{inputenc}
\usepackage[english]{babel}
\usepackage{times}
\usepackage{latexsym}
\usepackage{fancyhdr}
\usepackage{verbatim}
\usepackage{tabularx}
\usepackage{epsfig}
\usepackage{amsmath}
\usepackage{amssymb}
\usepackage{graphicx}
\usepackage{wasysym}

\usepackage{yfonts}

\newcommand{\qed}{\hspace*{\fill}$\square$}

\newcommand{\be}{\begin{equation}}
\newcommand{\ee}{\end{equation}}

 \newcommand{\Z}{\mathbf{Z}}

 \newcommand{\half}{\frac 1 2}

 \newcommand{\ket}[1]{|#1\rangle}

\begin{document}

\title{An Exclusion Principle for Sum-Free Quantum Numbers}

\author{Miguel A. Martin-Delgado}
\affiliation{Departamento de F\'{\i}sica Te\'orica, Universidad Complutense, 28040 Madrid, Spain.\\
CCS-Center for Computational Simulation, Campus de Montegancedo UPM, 28660 Boadilla del Monte, Madrid, Spain.}

\begin{abstract} 
A hypothetical exclusion principle for quantum particles is introduced that generalizes the  exclusion and inclusion principles for fermions and bosons, respectively: the correlated exclusion principle. The sum-free condition for Schur numbers can be read off as a form of exclusion principle for quantum particles. Consequences of this interpretation are analysed within the framework of quantum many-body systems. 
A particular instance of the correlated exclusion principle can be solved explicitly yielding a sequence of quantum numbers that exhibits a fractal structure and is a relative of the Thue-Thurston sequence. The corresponding algebra of creation and annihilation operators can be identified in terms of commutation and anticommutation relations of a restricted version of the hard-core boson algebra.
\end{abstract}

\maketitle

\tableofcontents

\section{Introduction}
\label{sec:conclusions}

The collective behaviour of quantum strongly correlated systems is the source of surprising exotic phenomena \cite{jaitisi,strongly}.
Even non-interacting particles, when they are identical \cite{Pauli},  manifest collective behaviour \cite{Bose,Einstein,Fermi,Dirac}
beyond any classical analog. They behave as if there were an effective force among them, repulsive for fermions and attractive for bosons.
The collective behaviour of fermions is crucial for explaining the periodic table of chemical elements \cite{Griffiths}, solid state properties like conductors vs. insulators \cite{Kittel}, the stability of matter \cite{Ehrenfest,Dyson_Lenard,Dyson}, white dwarfs and neutron stars in astrophysics, etc. 
Similarly, the collective properties of bosons are responsible for unusual physics like the superfluidity of $\text{He}^4$ \cite{Kapitza,Allen_Misener,Landau}, BCS superconductivity \cite{Onnes,BCS}, Bose-Einstein condensates \cite{BEC1,BEC2}, etc.

The fascinating world of the quantum properties exhibited by fermions and bosons has spurred
the quest for alternative exclusion principles exemplified by new quantum statistics \cite{Haldane}. A successful example of this search are anyons.
Anyons are a surprise of flatland physics \cite{LM,Wilczek}. Whereas in the ordinary three-dimensional world the only possible quantum statistics for point-like particles are fermionic or bosonic \cite{RAC2016},
in two-dimensions a new world of possibilities opens up for exchanging identical particles by means of braiding operations instead of the common permutations in 3D spatial dimensions. Originally, anyons were conjectured quasiparticles that are admissible by the quantum mechanics in two-dimensional particles but detached from real physics. An initial step towards a physical picture of anyons was proposed by Wilczek who envisaged 
charged vortices as a candidate for anyons \cite{Wilczek}. This construction relies on the Aharonov-Bohm effect \cite{AB}. That charged vortex is a composite of a charged particle with a magnetic flux attached to it living in two dimensions with the magnetic field perpendicular to the plane. By looping around these charged vortices they pick up a phase factor in the wave function that can range from 0 (bosons) to $\pi$ (fermions), being anyons any of the intermediate 
phases. This example illustrates how interactions may change the original exclusion principle of electrons into anyons.

Interestingly enough, the theoretical construct of Wilczek turned out to be closer to reality than expected. In fact, the quantum Hall effect  \cite{QHE} is a physical realization of the quantum flatland. Specifically, in the fractional quantum Hall effect \cite{FQHE1,FQHE2} the extreme conditions of low temperature and high magnetic fields make the electrons in the planar sample undergo a transformation of their original properties such that the excitations of the system are no longer simple free electrons but quasiparticles behaving like anyons depending on the filling factor of the sample \cite{FQHE3}. 

The physics of a system of interacting anyons, like  charged vortices, are generically very difficult to study and there are problems regarding their collective behaviour that still remain open. 
The simplest example of exactly solvable model of anyons is the topological code of Kitaev \cite{Kitaev}. 
This is another example of how particle interactions (in this case a type of geometrical interaction) can turn excitations of standard particles (localized spins) into quasiparticles with anyonic properties.
In this model, the resulting anyons are abelian anyons called semions for their effective flux is $\frac{\pi}{2}$, half way between bosons and fermions. The exact solvability 
facilitates the study of its thermal properties \cite{qutrits,thermal}. Moreover, the original two-dimensional barrier for obtaining quasiparticles with fractional statistics has been overcome and it is possible to realize them in three dimensions or arbitrary dimensions provided the excitations are extended objects like membranes or generalizations thereof called branyons \cite{3DTQC,Branyons}.

There has been an intense and long search for the experimental direct realization of anyons that still continues. The latest best evidence for them has been recently reported \cite{anyons2020} in a system of electrons in a nanostructured interferometer made with  gallium arsenide semiconductors.
The braiding quantum phase of these anyons is $\frac{2\pi}{3}$. What is at stake is the direct observation of a quantum phase with a braiding origin. Other manifestations of anyons have been reported with Majorana quasiparticles \cite{Majorana1,Majorana2,Majorana3} 
as well as attempts to using their braiding properties implicitly \cite{Litinski}.

On the contrary, there are alternative quantum statistics that have not yet met with experimental success, like parafermions and their parastatistics. These were introduced to solve an experimental puzzle in the emerging
theory of hadrons, the quark model \cite{quark}. The mysterious situation with the baryon $\Delta^{++}$ composed
of three up quarks with parallel spins was an indication of an apparent violation of the Pauli exclusion principle.
Parastatistics were postulated to solve this situation, but later the problem was resolved by postulating a new
$SU(3)$ gauge degree of freedom, later called the color charge carrying the fundamental force of QCD \cite{parafermions1,parafermions2}.

The correlated exclusion principle proposed here is another theoretical construct looking for a possible physical realization. The aim of this work is to investigate to some extent the consequences of it for a collection of quantum particles abiding by that hypothetical principle. 
The appellative correlated in the correlated exclusion principle precisely appeals to some sort of interaction that dresses the original particles into new interaction-free quasiparticles  but with a novel exclusion principle. This is in the same spirit as for the charged vortices \cite{Wilczek}, the Kitaev model \cite{Kitaev} or the quasiparticles in the topological color codes \cite{ColorCodes}.

An intuition as to why the calculation of Schur numbers involves such a high complexity is to notice that for a quantum system, the Schur numbers in the spectrum are the antithesis of the quantum harmonic oscillator.
The latter satisfies that given two energies $E_1$ and $E_2$ in the spectrum, the sum $E_1 + E_2$ is also allowed. Actually, the quantum harmonic oscillator is the only quantum system with this property. It is because of this feature that it is used in the formalism of second quantisation in many-body problems. Therefore, a sum-free condition in the spectrum as in the correlated exclusion principle is a signature of a strongly correlated many-body system at a very strong level of correlation.

In the real exclusion principle of Pauli, indistinguishable particles like fermions are subject to an effective repulsive force
as a consequence of the quantum degeneracy of fermionic quantum states. In the case of the quasiparticles subject to a correlated exclusion principle,
they have to interact through a real force that could possibly cause the existence of quantum states with the sum-free property of Schur numbers. Interestingly, quantum states with the sum-free property give rise to a intrinsic degeneracy (see Sec. \ref{sec:states}) that is a novel feature of the postulated exclusion principle.

\section{Schur Numbers}
\label{sec:numbers}

The Schur numbers were introduced by Issai Schur in 1916 \cite{Schur} in his work on the modular version of the Fermat's last theorem.
Schur numbers arise when studying sets of integer numbers with the property of 'sum-free sets'. This property amounts to have a given set of distinct integers $S$ such that the sum of sets $S+S$ does not contain the original elements of $S$. This is clarified with an example.
Let $S$ be $S=\{1,2\}$. This set is not sum-free since $S+S = \{2,3,4\}$ contains the 2 that is originally in $S$. However,
$S=\{1,3\}$ is sum-free since $S+S =  \{2,4,6\}$ does not contain elements of $S$.

\noindent {\em Definition 1}. Given an integer $K$, the Schur number denoted as $S(K)$ is defined as the maximum integer $n$ such that the integers $1,2,\ldots ,n$ can be distributed into $K$ partitions in such a way that the $K$ partitions are sum-free sets.

An important property is that Schur numbers exist and are finite for a given $K$. This is a consequence of a theorem by Schur.

\noindent {\em Theorem 1}. For any partition of the positive integers into a finite number of parts $K$, one of the parts contains three integers $n_1, n_2, n_3$ with $n_1+n_2=n_3$.

\noindent In his original proof, Schur provided an upper bound for the maximum positive integer $S(K)<\lfloor K! e \rfloor$. 

\noindent Denote the partitions of a set of distinct integers $S$ as $P=\{S_1,S_2,\ldots,S_k, \ldots, S_K\}$.
Let us see the simplest non-trivial example $S(2)$. Say we have to boxes (partitions) $S_1$ and $S_2$, or they can be labeled by colors as well. We start
placing the 1 in any of them, for instance, the $S_1$. 
\begin{equation}
\begin{aligned}
S_1 &= \{1\} \\
S_2 &=\{\}.
\end{aligned}
\end{equation}
Next, the 2 can only be placed at box $S_2$ for in the $S_1$ it collides with the case $1+1$. Thus,
\begin{equation}
\begin{aligned}
S_1 &= \{1\} \\
S_2 &=\{2\}.
\end{aligned}
\end{equation}
Now we arrive at 3 that can be placed in any of the boxes, but careful: if we place it in $S_1$, then the 4 can not go into $S_1$ for it collides
with $1+3$, and neither 4 can go into $S_2$ since it collides with $2+2$. However, if we place 3 in $S_1$ we have $S(2)=3$, but this would not be the
maximum partition possible into two boxes. There exists the possibility of placing 3 in $S_2$ and 4 in $S_1$. Thus, we have sum-free sets:
\begin{equation}
\begin{aligned}
S_1 &= \{1,4\} \\
S_2 &=\{2,3\}.
\end{aligned}
\end{equation}
This is the maximum partition we can have for if we try to accommodate the 5, we see that it is no longer possible since it collides in $S_1$ with $1+4$
and in $S_2$ with $2+3$. Then, we finally have $S(2)=4$.

\noindent Schur numbers exhibit a hard complexity and very few of them are known (see Table \ref{table_schurs}).
There are bounds to the growth of $S(K)$ \cite{Weisstein}. Schur proved \cite{Schur} that 
\begin{equation}
S(K+1) \geq 3 S(K) +1,
\end{equation}
from which it follows that
\begin{equation}\label{exponential}
S(K) \geq \half (3^K-1),
\end{equation}
and the equality is achived for $K=1,2,3$. A more refined lower bound is,
\begin{equation}\label{lowerbound}
S(K) \geq c \ 321^{K/5} > c \ 3.15977^K,
\end{equation}
for $K>5$ and $c$ a constant. Due to the strong complexity of Schur numbers $S(K)$, only $S(3)=13$, $S(4)=44$ and recently $S(5)=160$ are known (see Table \ref{table_schurs}).
\begin{table}    
    \begin{tabular}{ |c| c | c |c|c|c|c|c|}
    \hline \hline
    $K$      &  1 & 2 &  3 & 4  & 5  & 6 & 7 \\ \hline 
    $S(K)$ &  1& 4   & 13   & 44  &  160 & $\geq  536$ &   $\geq  1680$  \\ 
    \hline\hline
    \end{tabular} 
\caption{The known values or lower bound of Schur numbers $S(K)$ for low values of partitions $K$ \cite{bound_sevilla}. $S(4)$ was obtained by Golomb and  Baumert \cite{Baumert}, $S(5)$ by Heule \cite{Heule}.
The lower bounds for $S(6)$ and $S(7)$ are due to Fredricksen and Sweet \cite{Fredricksen,Sanz}.
} 
\label{table_schurs}
\end{table}

\noindent The partitions for $S(3)$ read as follows \cite{Kisner}:
\begin{equation}
\begin{aligned}\label{S(3)_1}
S_1 &= \{1,4,7,10,13\} \\
S_2 &=\{2,3,11,12\}\\
S_3 &=\{5,6,8,9\}.
\end{aligned}
\end{equation}
\begin{equation}\label{S(3)_2}
\begin{aligned}
S_1 &= \{1,4,10,13\} \\
S_2 &=\{2,3,7,11,12\}\\
S_3 &=\{5,6,8,9\}.
\end{aligned}
\end{equation}
\begin{equation}\label{S(3)_3}
\begin{aligned}
S_1 &= \{1,4,10,13\} \\
S_2 &=\{2,3,11,12\}\\
S_3 &=\{5,6,7,8,9\}.
\end{aligned}
\end{equation}
The partitions are three-times degenerate depending on whether the 7 is placed on each box. As for $S(4)$, here is one example of partition \cite{Baumert}:
\begin{equation}
\begin{aligned}
S_1&= \{1,3,5,15,17,19,26,28,40,42,44\} \\
S_2&=\{2,7,8,18,21,24,27,33,37,38,43\}\\
S_3&=\{4,6,13,20,22,23,25,30,32,39,41\}\\
S_4&=\{9,10,11,12,14,16,29,31,34,35,36\}.
\end{aligned}
\end{equation}
For an instance of partition for $S(5)$, see App. \ref{sec:Heule}.

\section{Quantum States}
\label{sec:states}
We can adscribe quantum states to the partitions associated to the Schur numbers in the following way. 
Let us denote by $\ket{n_1,n_2,n_3, \ldots }_{S_k} := \ket{n_1}_{S_k}\ket{n_2}_{S_k}\ket{n_3}_{S_k} \cdots $ a quantum state associated to the partition $S_k$ of the Schur number $S(K)$
such that $\ket{n_i}_{S_k}=:\ket{n_{i,k}}_{k}$ is a state associated to a certain quantum number $n_{i,k}$ representing the value of a physical observable of the i-th particle in the state of the partition $S_k$. The particular meaning of this quantum number will depend
on the particular physical realization. By construction, these quantum states will enforce the sum-free property for the quantum numbers from which they are constructed: given $n_{i,k}, n_{j,k} \in S_k$ then $n_{i,k} + n_{j,k} \not\in S_k$.
For example, for the partitions of $S(3)$ \eqref{S(3)_1},\eqref{S(3)_2},\eqref{S(3)_3} we may write:
\begin{equation}\label{S(2)}
\ket{\Psi(1)}= \ket{1,4,7,10,13}_{1}
\ket{2,3,11,12}_{2} \ket{5,6,8,9}_{3}.
\end{equation}
\begin{equation}
\ket{\Psi(2)}= \ket{1,4,10,13}_{1}
\ket{2,3,7,11,12}_{2} \ket{5,6,8,9}_{3}.
\end{equation}
\begin{equation}
\ket{\Psi(3)}= \ket{1,4,10,13}_{1}
\ket{2,3,11,12}_{2} \ket{5,6,7,8,9}_{3}.
\end{equation}

We may envisage that a quantum superposition of those states is possible, assuming there is no superselection rule forbidding it:
\begin{equation}\label{S(3)}
\ket{\Psi(S(3))}=  \frac{1}{\sqrt{3}} (\ket{\Psi{(1})}_{123} + 
\ket{\Psi{(2)}}_{123}  + \ket{\Psi{(3)}}_{123}).
\end{equation}
Although the order of the numbers in the partitions $S_k$ of a given $S(K)$ is arbitrary once it is constructed, this is not the case in the quantum state
corresponding to the partition  $S_k$ since the tensor product has a fixed order. To this end, we shall define the tensor product of quantum states comprising the partition $S_k$ as ordered in increasing values of the positive integers of $S_k$. Thus, despite dealing with non-identical particles,
it is possible to have a source of degenenacy of quantum states that is intrinsic to the sum-free condition of the quantum numbers.

There is a clear difference between the quantum states  \eqref{S(2)} and \eqref{S(3)} under the assumption that the sum-free condition
applies for the quantum numbers involved: whereas \eqref{S(2)}  is a product state, \eqref{S(3)} is a quantum supperposition. This fact shows
that the quantum numbers in \eqref{S(2)} are blocked and cannot change from one state to another within it. On the contrary, \eqref{S(3)} shows
the novel feature that one quantum state, $\ket{7}$, can be shared by any of the three coherent components of the total state \eqref{S(3)}, while
the remaining quantum states are blocked. A physical interpretation of this effect is shown in Sec. \ref{sec:correlated}. Since the value 7 always occurs for the third particle for the three partitions of $K=3$, the superposition state \eqref{S(3)} is factorizable as:
\begin{equation}\label{S(3)bis}
\begin{aligned}
\ket{\Psi(S(3))}&= \ket{1,4,10,13}_{1}\ket{2,3,11,12}_{2} \ket{5,6,8,9}_{3}  \\
                        &\frac{1}{\sqrt{3}} (\ket{7_3}_{S_1} + 
\ket{7_3}_{S_2}  + \ket{7_3}_{S_3}).
\end{aligned}
\end{equation}
Thus, an interesting question is wether there are entangled states for higher order partitions:

\noindent {\em Unsolved}: Determine the existence of quantum entangled states for patitions $K\geq 4$.

\begin{figure}[t]
  \includegraphics[width=0.5\textwidth]{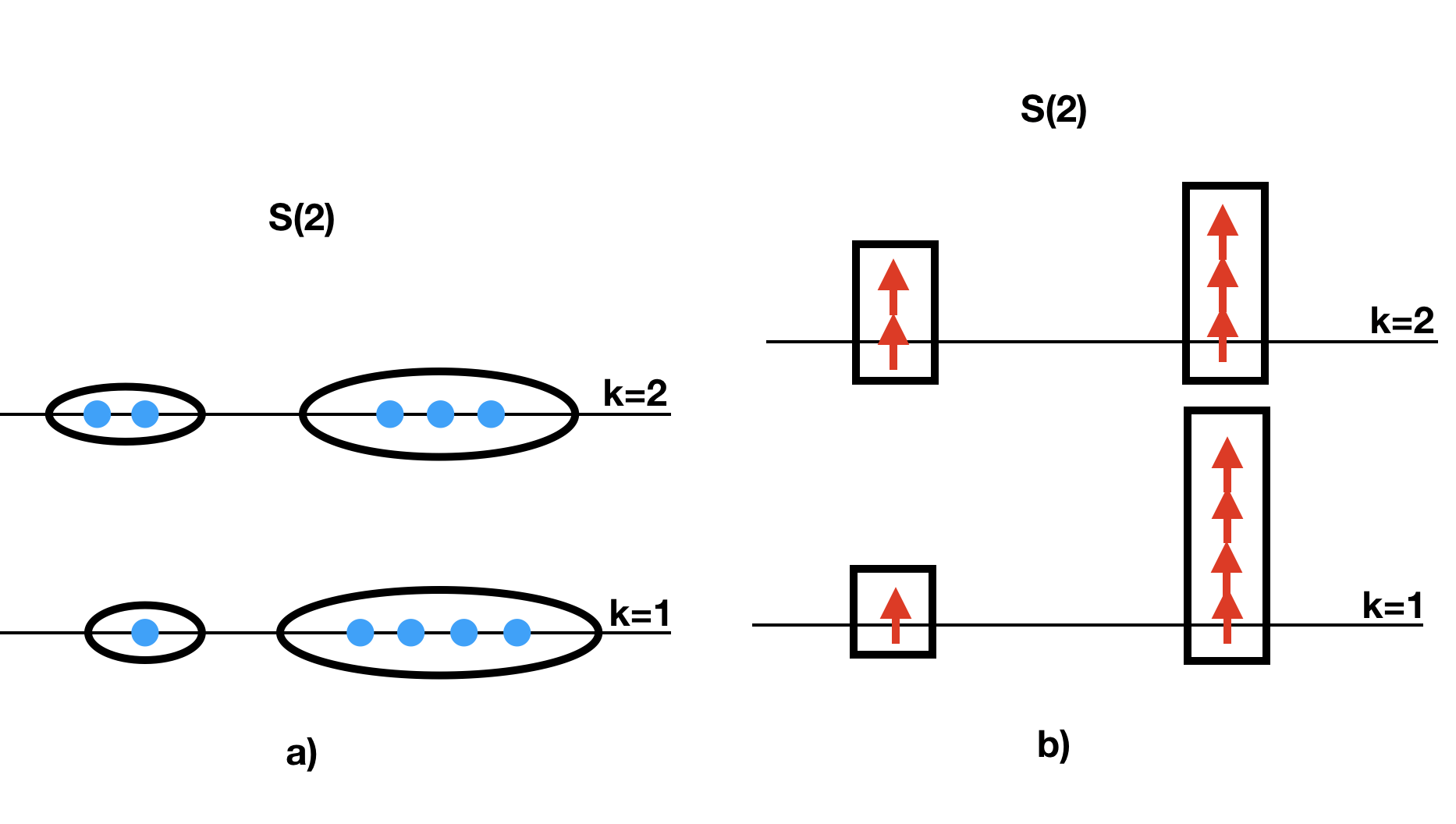}
  \caption{Some possible physical pictures of quasiparticles obeying the exclusion principle for sum-free quantum numbers: a) bosons; b) spins (see explanations in text).}
  \label{fig:physical}
\end{figure}

It is instructive to give some physical picture to these quantum states. For instance, the different partitions $S_k, 1\leq k \leq K$ can be thought of as energy levels. This is not the only possibility and realizations in terms of set of sites on a lattice or other interpretations may work as well.  Let us list some possibilities.

\noindent {\bf Bosons.} We may think of the values $n_{i,k}$ as the number of $n_i$ bosons forming individual quasiparticle groups placed at a certain energy level $k$, see Fig. \ref{fig:physical}.  The number of bosons is an additive physical quantity. 

\noindent {\bf Spins.} In this case, the values $n_{i,k}$ represent the number of elementary spins polarized in the same direction, say up, see Fig. \ref{fig:physical}. In particular, this includes the possibility of fermions making up the quasiparticles.

\noindent {\bf Qudits.} The values $n_{i,k}$ may correspond to a set of multilevels of a qudit  system (see also Sec.\ref{sec:modular}). 
For example, the multilevels may represent $2s_{i,k}+1$ spin levels, or harmonic oscillator levels. If the value  $n_{i,k}$ is even, the spin is integer; whereas if it is odd, the spin is half-integer. The third component of the spin is also additive.

\begin{figure}[t]
  \includegraphics[width=0.35\textwidth]{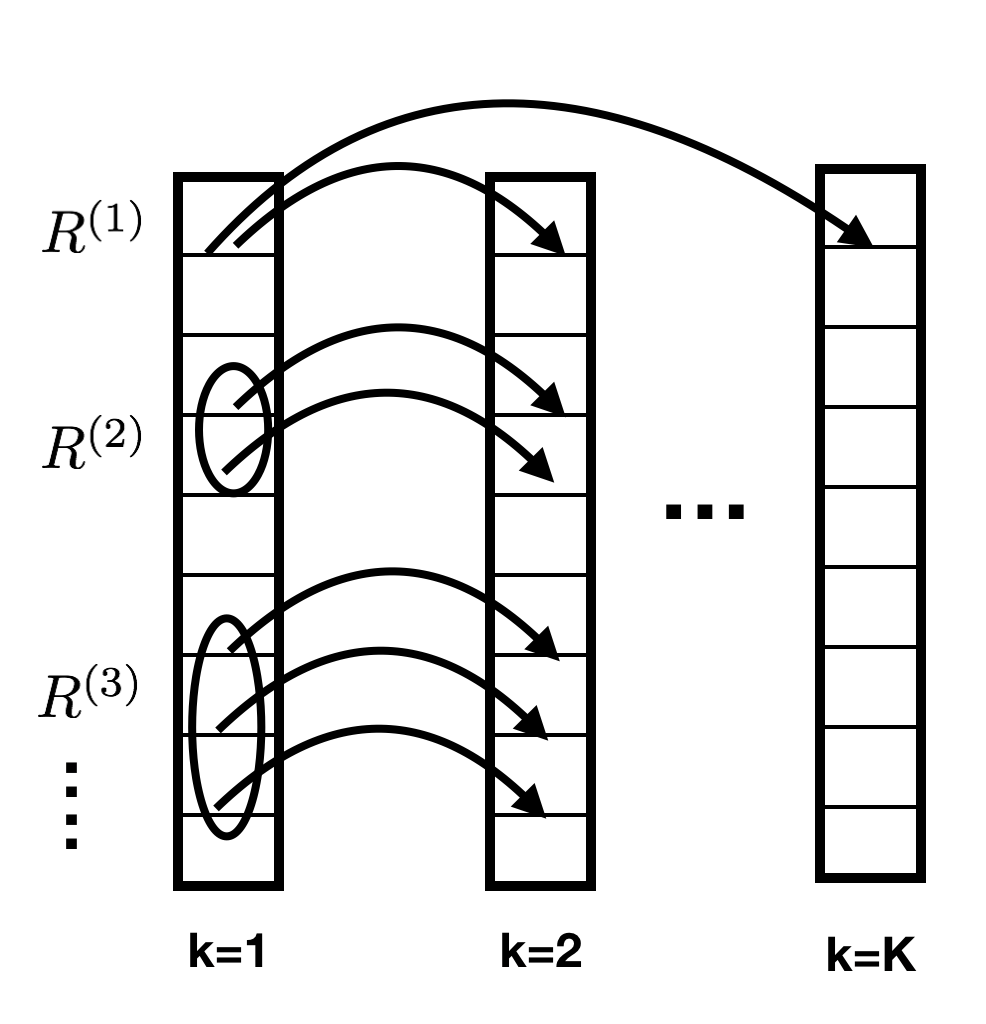}
  \caption{A pictorial view of the transformations forming the set ${\cal R}$ of invariant transformations leaving invariant the partition functions $S_k, 1\leq k \leq K$  (see explanations in text).}
  \label{fig:transformations}
\end{figure}

\noindent What remains unknown is the physical origin of the sum-free rule in all these cases.

Let ${\cal R}$ be the set of transformations that leave invariant the $K$ levels associated to the quantum states of $S(K)$.
An element $R\in {\cal R}$ is a rearrangement of the quantum numbers $n_{i,k}$ for each level $S_k, 1\leq k \leq K$ such that the
new quantum states for quantum numbers  $n^\prime_{i,k}$ satisfy the sum-free rule. Explicitly, given $n_{i,k}$, sum-free $n_{i,k}+ n_{j,k} \notin S_k$, then
\begin{equation}\label{set}
R \ket{\Psi_k(n_{1,k},n_{2,k},\ldots)} = \ket{\Psi_k(n^\prime_{1,k},n^\prime_{2,k},\ldots)},
\end{equation}
with $n^\prime_{i,k}$ sum-free, $n^\prime_{i,k}+ n^\prime_{j,k} \notin S_k$. Then, the cardinality $|{\cal R}|$ of ${\cal R}$ is the degeneracy of the quantum states associated to $S(K)$. This degeneracy is intrinsic to the sum-free states. Permutations are not allowed transformations since particles are treated as non-identical, and thus distinguishable.

Knowing  ${\cal R}$ is a complex task since it relies on the determination of the partitions of $S(K)$. However, if we are given an instance partition of $S(K)$, then there is an efficient procedure to generate all the elements of ${\cal R}$. This is done as follows.

\noindent {\bf First Order Transformations $R^{(1)}$.}
Assume that each quantum state $\ket{\Psi_k(n_{1,k},n_{2,k},\ldots)}$ is labeled with values in increasing order 
${n_{1,k} < n_{2,k} < \ldots}$. Take the first quantum state corresponding to $k=1$ as a reference state. Take the greatest value  $n_{\text{max},1}$. Consider the rest of quantum states $k=2,3,\ldots K$, see Fig. \ref{fig:transformations}. By removing this value number from the state $k=1$, it remains a sum-free state. Let us insert $n_{\text{max},1}$ in the state $k=2$ at the greatest position compatible with the increasing order of the quantum numbers. First thing we have to check is that the value $2n_{\text{max},1}$ is not in the state $k=2$.
Otherwise, the procedure stops. If that condition is satisfied, we check for extra conditions to make the new $k=2$ state sum-free. This can be done efficiently by substracting from $n_{\text{max},1}$ the values of $k=2$ smaller than $n_{\text{max},1}$ and checking wether the difference is not in $k=2$. If this condition is also satisfied, then the new state denoted as $k^\prime = 2$ is an acceptable sum-free state and we have obtained a new partition compatible with $S(K)$. The corresponding transformation of quantum numbers is an element of  ${\cal R}$. This procedure can be repeated for all the quantum numbers in each of the quantum states $k=2,3,\ldots, K$.
An example of this type of transformation for $S(3)$ is given in eqs. \eqref{S(3)_1}, \eqref{S(3)_2}, \eqref{S(3)_3} where the elements amount to permutation of the value 7 among the $K=3$ quantum states. These type of transformations involving the rearrangement of one occupation number will be called transformation of first oder, denoted as $R^{(1)}$. In particular, for $S(3)$ there are only first order transformations and the set  
${\cal R}$ contains 3 elements:
\begin{equation}
{\cal R} = \{ 1, R_1^{(1)},R_2^{(1)}\},
\end{equation}
defined as
\begin{equation}
\begin{aligned}
R_1^{(1)}\ket{1,4,7,10,13}_{S_1} &:= \ket{2,3,7,11,12}_{S_2},\\
R_2^{(1)}\ket{1,4,7,10,13}_{S_1}&:= \ket{5,6,7,8,9}_{S_3}.
\end{aligned} 
\end{equation}

\noindent {\bf Second Order Transformations $R^{(2)}$.}
Similarly, transformations of second order $R^{(2)}$ involve the rearrangement of two quantum numbers from state $k=1$ into
another single state from the set $2\leq k \leq K$. We do not need to consider the case of inserting two values from the state $k=1$ into two different states, say $k=2,3$ since they correspond to the composition of first order transformations. 

\noindent {\bf Higher Order Transformations $R^{(l)}, l=3,\ldots$.}
And so on and so forth for higher order transformations.

\begin{figure*}
  \includegraphics[width=0.6\textwidth]{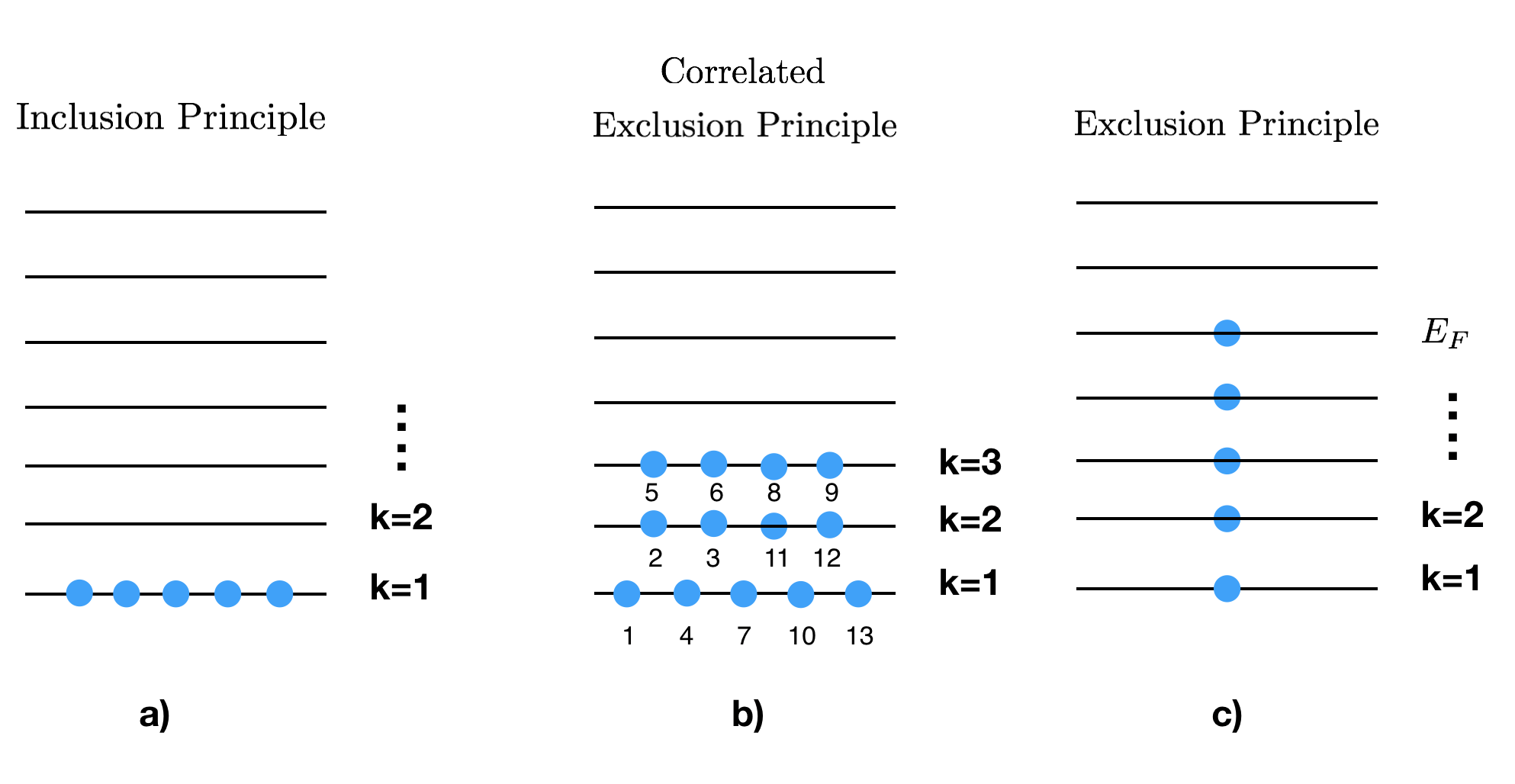}
  \caption{A pictorial view of the ground state for a) the inclusion principle, b) the correlated exclusion principle and c) the exclusion principle.}
  \label{fig:ground_state}
\end{figure*}

However, there is a caveat regarding the composition law of two of such transformation operations. Denote by $\tilde{R}^{(1)}$
a first order transformation that is not in ${\cal R}$ for it does not satisfy the sum-free rule. Yet, by composing two such a not allowed transformations, we can arrive at
\begin{equation}
\tilde{R}_1^{(1)} \tilde{R}_2^{(1)}:= R_{12}^{(2)},
\end{equation}
which in turn is an allowed transformation of second order $R_{12}^{(2)}\in {\cal R}$. An example of this is an exchange transformation between states $k=1$ and $k=2$.
Similarly, given two allowed transformations of 
first order we may compose them obtaining a non-allowed transformation,
\begin{equation}
R_1^{(1)} R_2^{(1)}:= \tilde{R}_{12}^{(2)}.
\end{equation}
Then we arrive at the following conclusion: {\em  the set ${\cal R}$ of invariant transformations of $S(k)$ is not guaranteed to be a group for the sum-free property is not satisfied by the group composition law}.

\noindent For the particular case of $S(3)$, ${\cal R}$ happens to be a group: the permutation group of the value 7 among the $K=3$ levels.
It is a $\Z_3$ cyclic group,
\begin{equation}
{\cal R}_{S(3)} = \Z_3,
\end{equation}
whereas for $S(1)=1$ and $S(2)=4$, it is simply the trivial group of a single element $\{1\}$.

\noindent {\em Unsolved}: Determine the set ${\cal R}_{S(4)}$ and whether it is a group.

\section{Correlated Exclusion Principle}
\label{sec:correlated}
Let us recall for a reference the basic principles describing systems of identical but indistinguishable particles
in quantum mechanics. 

\noindent {\bf Exclusion Principle:}
Two fermions can not occupy the same quantum state.

\noindent {\bf Inclusion Principle:}
Any number of bosons can occupy the same quantum state.

A quantum state is determined by a set of quantum numbers representing 
physical observables like position, linear momentum, spin etc. Thus, bosons 
and fermions behave in opposite manner with respect to how the energy levels
of the spectrum are filled to form the ground state of a quantum system at zero temperature.
The exclusion principle was introduced by W. Pauli \cite{Pauli} in his studies of multielectronic atoms.
The inclusion principle originates from the works of S. Bose  \cite{Bose} and A. Einstein  \cite{Einstein}  about the collective
behaviour of photons and atoms.
In order to make sense of those principles, they have to be supplemented with an operational
characterization of what fermions and bosons are. This is provided by the Spin-Statistics theorem \cite{Pauli2,PCT}
that relates bosons with particles with integer spin and fermions with half-integer spin. In the modern
theory of identical quantum particles, these principles are stated in terms of the total symmetrization or antisymmetrization
of the wave function with respect to permutations of the quantum numbers of the individual particles.
However,  we prefer to keep the older version of these principles since they fit better into the current discussion.

Now, we can construct an exclusion principle
for non-identical particles associated to the Schur quantum states of Sec.\ref{sec:states}. The fact that
the quantum numbers $n_{i,k}$ entering those states are different one another makes
those particles intrinsically non-identical, and thus distinguishable, whatever the particular physical realization
of those numbers.  
We propose to interpret the sum-free condition on quantum states as a sort of exclusion principle:
the one that forbids particles with quantum numbers in the partition $S_k$ to appear in the set $S_k+S_k$,
$1\leq k \leq K$. A possible realization of the different partitions $S_k$ of a given $S(K)$ Schur number is 
associated to $K$ energy levels, each with energy $E_k$. Denote those energy subspaces as ${\cal E}_k$ with
$k=,1,2,\ldots,K$ and we can identify the abstract sets $S_k$ with energy levels ${\cal E}_k$, but we may think of other
possible realizations in coordinate space (e.g. one-dimensional lattices) or else, as appropriate.

\noindent {\bf Correlated Exclusion Principle:}
When two particles occupy states with quantum numbers $n_{i,k}$ and $n_{j,k}$ in the spectrum ${\cal E}_k$, then a third particle can not occupy the  quantum state corresponding to $n_{i,k}+n_{j,k} \in {\cal E}_k$.

The interpretation of the sum-free condition for sets $S_k$ of quantum numbers is very natural and generalizes the conditions
represented by the real exclusion and inclusion principles. We can see this with the process of filling up the energy levels  ${\cal E}_k$ according to the correlated
exclusion principle. This was done in Sec. \ref{sec:states} for low numbers of energy levels $K=1,2,3,4$. The most prominent features of the exclusion/inclusion principles manifest in the structure of the ground state of the quantum-many body system. It is very illustrative to compare it with the estructure that emerges for the correlated exclusion principle as shown in Fig.\ref{fig:ground_state}. The procedure to obtain the ground state for each principle is to start filling up the energy levels according to the rules dictated by each principle. When the filling is done, the emerging structure leaps to the eye. The inclusion principle is represented by the Bose-Einstein condensate and the exclusion principle by the Fermi energy for the tower of fermions. The ground state of a correlated exclusion principle sits in the middle of this extreme conditions: it shares with the exclusion principle the fact that no particle is allowed to be repeated with the same energy and it shares with the inclusion principle the possibility of having several particles at the same energy level. It is clear that the new ground state relies on distinguishable (non-identical) particles.

In this regard, the correlated exclusion principle is related to the parafermions, which also exhibit intermediate properties between bosons and fermions. Their ground state is described by Young tableaux dictating hybrid symmetrization and anti-symmetrization rules for the wave function \cite{GP}.

Now it is possible to give another more physical view to the quantum states that arise from the Schur number $S(3)$ in \eqref{S(3)_1},  \eqref{S(3)_2} and  \eqref{S(3)_3}. With the energy level interpretation of the partitions $S_1,S_2,S_3$ in Fig.\ref{fig:ground_state}, the case  \eqref{S(3)_1} represents the lowest energy state, whereas the other two states have higher energy corresponding to placing the particle with quantum number 7 in higher energy levels. Looking at these states \eqref{S(3)_1},  \eqref{S(3)_2} and  \eqref{S(3)_3} as if they were electronic configurations of atoms, we see that the particle 7 can be not only in the higher energy level, but also in the other lower energy levels that are blocked by the correlated exclusion principle and thus would correspond to full valence shells in the atomic language.

The absence of a known topological origin for the correlated exclusion principle puts it on a different footing than the exclusion/inclusion principles for identical non-interacting particles. Thus, it is more natural to think that some sort of dynamics could be responsible for the interactions built upon free particles so that the resulting dressed particles by the interactions may comply with the sum-free condition of quantum numbers. In fact, it is worthwhile to mention that the quantum spins comprising topological models, like the quantum error correction codes of Kitaev \cite{Kitaev} and topological color codes \cite{ColorCodes}, are dressed by peculiar geometrical interactions producing emergent quasiparticles that, despite being non-identical, they exhibit non-trivial anyonic braiding statistics.

So far for the zero-temperature properties of the ground state wave function of quantum particles obeying the correlated exclusion principle. As for finite-temperature effects, we need to have the guiding principles governing the construction of quantum states for higher number of partitions $K$, which is absent for now.

\section{Creation and Annihilation Operators}
\label{sec:operators}

We may construct these quantum states within a second quantization formalism 
by introducing creation and annihilation operators $B_{n_{i,k}}$, $B^\dagger_{n_{i,k}}$ and the 
corresponding vacuum state $\ket{\varnothing}$. In this context, it is important to notice that the symbol
$\varnothing$ is different from 0: the latter means that the physical property representing the quantum
numbers $n_{i,k}$ takes on the value 0, as for example a zero spin particle, whereas the symbol $\varnothing$
represents the absence of a particle, it is literally nothing. It is because of this distinction that we can
work with states in second quantization without running into the contradiction that 0 is incompatible with
the property of sum-free states since $0+0=0$. 

There are at least two possible ways to realize these quantum states:

1. Statistics: To assume that the particles are non-interacting and that the arrangements of occupation numbers in the quantum states
come from some form of quantum statistics. Examples of these are bosons and fermions.

2. Interactions: To assume that the particles are normal particles and some form of interactions are responsible for the pattern of occupation numbers.

Let $\ket{n_{1,k},n_{2,k},\ldots}_k$ be a quantum state corresponding to a certain partition $S_k,1\leq k \leq K$ with a number of particles $S(K)$ that 
are distributed according to the the sum-free property of $n_{i,k}$. These states are obtained from the vacuum states $\ket{\varnothing,\varnothing,\ldots }$ upon acting with the creation operators,
\begin{equation}
\ket{n_{1,k},n_{2,k},\ldots}_k:=  B^\dagger_{n_{1,k}} B^\dagger_{n_{2,k}} \cdots \ket{\varnothing,\varnothing,\ldots }.
\end{equation}
The vacuum state is defined through the annihilation operators:
\begin{equation}
 B_{n_{i,k}} \ket{\varnothing,\varnothing,\ldots } = 0, \forall i,k.
\end{equation}

To properly account with these creation/annihilation operators, we must specify some algebraic structure similar to canonical commutation or anti-commutation relations. Annihilation operators can be declared commuting operators since removing a particle from $\ket{n_{1,k},n_{2,k},\ldots}_k$ with the sum-free property of $n_{i,k}$ does not contradict the sum-free property. Thus,
\begin{equation}
[B_{n_{i,k}},B_{n_{j,k}}] = 0 \ \forall i,j.
\end{equation}
However, the creation operators $B^\dagger_{n_{i,k}}$ clearly cannot be commuting operators since there are constraints that they have to fulfill in order to preserve the sum-free property. If we create a particle first with 
$B^\dagger_{n_{i,k}}$, then a particle created next with $B^\dagger_{n_{j,k}}$ must comply with the sum-free condition, 
\begin{equation}
B^\dagger_{n_{i,k}} B^\dagger_{n_{j,k}} \ket{\ldots,n_{i,k},n_{j,k},\ldots}_k =
 \begin{cases}
0 & \text{if}\  n_i + n_j \in S_k \\
 \neq 0 & \text{if}\  n_i + n_j \notin S_k 
\end{cases}
\end{equation}
Thus, the order in which particles are created does crucially matter. In particular, if we focus on the single property $n_{i,k}=n_{j,k}$
we arrive at an exclusion property:
\begin{equation}
(B^\dagger_{n_{i,k}})^2\ket{\ldots,n_{i,k},\ldots}_k =0.
\end{equation}
This is a consequence of the sum-free property since once we have a particle with the property $n_{i,k}$ occupying a state, we cannot place another one with the same property for then  $2n_{i,k}\in S_k$.  We shall get back to this in the next section, but this condition suggests an anticommutation relation for the creation operators at the same state:
\begin{equation}
\{B^\dagger_{n_{i,k}},B^\dagger_{n_{i,k}}\} = 0.
\end{equation}

The problem is that we do not now a priori
these ordering rules since for that we would have to solve the problem of the Schur partitions and numbers.
We may describe this situation with an anticommutation rule in a rather generic form:
\begin{equation}\label{g}
\{B^\dagger_{n_{i,k}},B^\dagger_{n_{j,k}}\} = g(N_{n_{i,k}},N_{n_{j,k}}) (1-\delta_{n_{i,k},n_{j,k}}).
\end{equation}
where $g$ is an unknown function of the number operators $N_{n_{i,k}}$. Notice that the choice of anticommutation relations for the creation operators is made in order to be compatible with the commutation relations of the annihilation operators. This would not be possible if both satisfy commutation relations.

Similarly, creation and annihilation operators cannot commute for the same reason as before due to the presence of creation operators, resulting in a commutation rule as
\begin{equation}\label{f}
\{B_{n_{i,k}},B^\dagger_{n_{j,k}}\} = f(N_{n_{i,k}},N_{n_{j,k}}).
\end{equation}
with $f$ some appropriate function. In this case, we still have the freedom to choose commutation relations as well.

A consequence of these commutation/anticommutation relations is that they do not close a linear algebra. The presence of number operators makes them intrinsically non-linear. However, this is not the first time that algebras like these appear.
A prominent example is the case of hard-core bosons \cite{LSM,Rigol} for which the algebra is also non-linear:
\begin{equation}
[b_{i},b^\dagger_{j}] = (1-2n)\delta_{i,j}.
\end{equation}

It is customary to name the quasiparticles or excitations associated to the creation/annihilation operators. In our case, we notice that
given two values of the sum-free property $n_{i,k}$ and $n_{j,k}$, they become allowed or coupled 
whenever its sum is not allowed to be present as an additional state. We may think of them as a 'quantum duet'. Thus, the states at levels $S_k$ are constituted by allowed duets. Following the tradition of naming particles with the end on, we shall call them {\em duetons}, for the lack of a better name.

Since the algebra of duetons comprises both commutation and anticommutation relations, we may think of them as a novel form of hard-core bosons, but with clear differences. One is the fact that hard-core bosons are indistinguishable particles, while duetons are distinguishable (non-identical). In the next section we shall see more explicit differences.

Another possible instance where to realize the sum-free property of quantum numbers is by means of fusion rules. For instance,
the parafermions mentioned earlier satisfy fusion rules by means of vertex operator algebras \cite{fusionrules}. The idea would be to construct new fields by imposing the sum-free constraint on the quantum numbers of the fusion rules instead of commutation/anticommutation relations.

The dynamics for the particles associated to these creation/annihilation operators can be generated by Hamiltonian terms defined in the space of 
energy levels ${\cal E}_k$ corresponding to the partitions $S_k, k=1,2,\ldots,K$. An energy term is the following:
\begin{equation}
H_0 = \sum_{n_{i,k}} E(k) B^\dagger_{n_{i,k}} B_{n_{i,k}},
\end{equation}
where $E(k)$ is the energy of the level $k$. Excitation and de-excitation processess can be produced by a jumping term like,
\begin{equation}
H_e = J \sum_{n_{i,k}\neq n_{j,k^\prime}} B^\dagger_{n_{i,k}} B_{n_{j,k^\prime}} + h.c.
\end{equation}
Interaction between levels can be represented in a local way by means of terms like
\begin{equation}
H_i =  \lambda \sum_{n_{i,k},n_{j,k^\prime}} N_{n_{i,k}} N_{n_{j,k^\prime}}.
\end{equation}
The total Hamiltonian is,
\begin{equation}
H = H_0 + H_e + H_i.
\end{equation}
The goal behind a Hamiltonian like this is to have its ground state corresponding to the sum-free quantum states of Sec.\ref{sec:states} (see Fig.\ref{fig:ground_state}) for some value of the
coupling constants $J$ and $\lambda$. By construction, this is true for $J=\lambda=0$. This must be true for increasing values of the number of particles $S(K)$ and energy levels $K$.

\section{Selfcorrelated Exclusion Principle}
\label{sec:selfcorrelated}

A particular and simpler case of the sum-free property occurs when we only demand
that a given quantum number $n_{i,k}$ is correlated with itself but not with the other values forming duetons. 
This corresponds to the condition $2S\not\subset S$ in set language. This way, the complexity of 
finding the partitions is enormously simplified since now the property of being sum-free
values only depends on each quantum number alone and not on the rest.
However, this simplified condition comes with the absence of the intrinsic degeneracy of 
the sum-free states that is present in the more general correlated exclusion principle.
This implies that there is only one level $K=1$ and the number of different quantum particles is simply $S(1)=N$, with $N$ 
the total number of allowed values satisfying the sum-free rule $2S\not\subset S$.
The state of the single partition $K=1$ is given by filling up sequentially according to the sum-free rule:
\begin{equation}\label{fractal}
\begin{matrix}
\ket{1,3,4,5,7,9,11,12,13,15,16,17,19,20,\ldots \bullet_N}_{K=1}
\end{matrix}
\end{equation}
 The occupation numbers of this state correspond to the sequence of integers $n$ such that $2n$ is not in the sequence.
 This induces a peculiar filling pattern of the energy level. This sequence has been rediscovered in different ways \cite{Sloane,Allouche} 
 and it is a relative of the famous Thue-Thurston sequence since it encodes the length of the blocks of the latter. The quantum states can be defined recursively, from the initial state $n_1=1$,
 \begin{equation}\label{recursion}
n_{i+1} = \begin{cases}
n_i+1 &  \text{if} \ (n_i+1)/2\not\in S_1 \\
n_i+2& \text{otherwise}.
\end{cases}
\end{equation}
The sequence of quantum numbers in the quantum state \eqref{fractal} has a fractal or self-assembly property: notice that the sequence of odd numbers is a subset, then remove them. We obtain a sequence of even numbers,
\begin{equation}
\ket{4,12,16,20,28,36,44,48,52\ldots \bullet_N}_{K=1}
\end{equation}
that is divisible by 4, after which we are back to the original quantum state \eqref{fractal}. The generating function of the occupation numbers is,
\begin{equation}
\sum_{i\geq 0} n_i x^i = \frac{1}{1-x} \prod_{i\geq 1} (1+x^{e_i}),
\end{equation}
with exponents starting at $e_1=1$ and given by,
\begin{equation}
e_{i+1} = \begin{cases}
2e_{i} +1&  \text{if} \  i \ \text{is even}\\
2e_{i} -1 & \text{if} \ \  i \ \text{is odd}.
\end{cases}
\end{equation}

We can resolve the simple sum-free condition $2S\not\subset S$ by means of the recursion relation \eqref{recursion}
and to obtain an explicit expression for the algebra of creation and annihilation operators, in constrast to the general case of Sec.\ref{sec:operators}. First notice that
there is a double exclusion effect when filling the unique state with $K=1$ \eqref{fractal}. On the one hand, for each value 
of $n_i$ at position $i$, the quantum number $2n_i$ is forbidden. This implies that double occupancy is excluded, similar to a fermion. 
In terms of creation operators this means $(B^\dagger_{n_i})^2=0$, or with anticommutation relations
\begin{equation}
\{B_{n_i},B^\dagger_{n_i}\}=0, \ \forall i.
\end{equation}
On the other hand, the values $n_i$ can be annihilated freely. Thus, we can assume a commutation
relation for them,
\begin{equation}
[B_{n_i},B_{n_j}]=0,\ \forall i,j.
\end{equation}
Yet, when the values $n_i\neq n_j$ are different, there is another exclusion effect for certain values of $n_i$ that are not allowed by
the sum-free condition \eqref{recursion}. This will be reflected in the commutation relation $[B_{n_i},B^\dagger_{n_j}]$. Denote the action of 
$B_{n_i}$, $B^\dagger_{n_i}$ on the states $\ket{\ldots \varnothing_i \ldots}$ and $\ket{\ldots n_i \ldots}$ as follows,
\begin{equation}
B_{n_i}\ket{\ldots \varnothing_i \ldots}=0, B_{n_i}\ket{\ldots n_i \ldots}=\ket{\ldots \varnothing \ldots};
\end{equation}
\begin{equation}
B^\dagger_{n_i}\ket{\ldots \varnothing_i \ldots}=\ket{\ldots n_i \ldots}, B^\dagger_{n_i}\ket{\ldots n_i \ldots}=0.
\end{equation}
Let us introduce the number operator $N_{n_i}$ as the one counting how many particles with the quantum numbers $n_i$ exists at the location $i$,
namely,
\begin{equation}
N_{n_i}\ket{\ldots \varnothing_i \ldots}=0, N_{n_i} \ket{\ldots n_i \ldots} = \ket{\ldots n_i \ldots}.
\end{equation}
With these relations, the action of the commutator $[B_{n_i},B^\dagger_{n_i}]$ on the basis states is
\begin{equation}
\begin{matrix}
[B_{n_i},B^\dagger_{n_i}]\ket{\ldots \varnothing_i \ldots}&=+\ket{\ldots \varnothing_i \ldots},\\
[B_{n_i},B^\dagger_{n_i}] \ket{\ldots n_i \ldots} &= -\ket{\ldots n_i \ldots}.
\end{matrix}
\end{equation}
For different occupation numbers $n_i\neq n_j$, the operators commute since there is no restriction on the order in which they are created and annihilated, except for the exclusion condition that some values for $n_i$ even are not allowed. Except for this latter condition, the algebra is like a hard-core boson algebra with an extra constraint. This can be summarized in the following commutation relation,
\begin{equation}
[B_{n_i},B^\dagger_{n_j}] = \begin{cases}
\varnothing & \text{if} \ n_i, n_j \not\in S_1 \eqref{fractal} \\
(1-2N_{n_i})\delta_{i,j} & \text{otherwise}.
\end{cases}
\end{equation}
Therefore, for this particular case of sum-free condition, we have succeded in finding the algebra of the creation and annihilation operators representing a selfcorrelated exclusion principle. With this, we can identifiy the functions $f$ \eqref{f} and $g$ \eqref{g} in the algebra for the general case of duetons. The result is a peculiar case of the hard-core boson algebra. 

\section{Modular Schur Numbers}
\label{sec:modular}

Schur \cite{Schur} proved that Schur numbers are finite for a given number of partitions $K$.
Nevertheless, the resources needed to account for observables yielding quantum numbers $n_{i,k}$ 
are very large as the lower bound shows \eqref{lowerbound}. This is clearly a limitation for a practical realization
of a system like this, although there are systems with orbital angular momenta of light that provide large quantum numbers values \cite{OAM1,OAM2,OAM3}.
Thus, it is interesting to think of generalizations of Schur numbers that may include a cut-off while
preserving some of their properties. There are many generalizations of Schur numbers \cite{generalizations}.
A possible way out is provided by modular Schur numbers \cite{modular}. In this case, the sum-free condition for positive integer numbers is defined modulo $m$, for $m$ a given positive integer. Thus, 

 \noindent {\em Definition 2}.
A set $S^{(m)}$ is said to be sum-free modulo $m$ when  $\forall n_i, n_j \in S^{(m)}$, then 
 $n_i +  n_j = z \ mod \ m \not\in S^{(m)}$. 
 
 \noindent Then, $S^{(m)}$ is a modular sum-free set. Then the definition of modular Schur numbers modulo $m$ follows:
 
 \noindent {\em Definition 3}. Given an integer $K$, the modular Schur number denoted as $S_m(K)$ is defined as the maximum integer $n$ such that the integers $1,2,\ldots ,n$ can be distributed into $K$ partitions in such a way that the $K$ partitions are modular sum-free sets modulo $m$.

It is apparent that multilevel quantum systems provide a natural realization of modular sum-free quantum numbers. These are called qudits of dimension $D$ and defined over the ring $\Z_D:=\Z /D\Z$, which is a field when $D$ is a prime number \cite{rmp,Distillation}

Some properties of modular sum-free numbers follow straightfowardly. Since a modular sum-free condition is also a sum-free condition a fortiori, it follows
\begin{equation}
S_m(K) \leq S(K).
\end{equation}
In addition, as $m + \ldots + m \equiv m \ mod \ m$, a modular sum-free set does not contain the modulus $m$ itself. This imposes a severe constraint to modular Schur numbers:
\begin{equation}
S_m(K) \leq m-1.
\end{equation}
Thus, the exponential growth of Schur numbers \eqref{exponential} is traded off for a constant growth of modular Schur numbers with atmost a linear growth with respect to the modulus $m$. The exact values of the modular Schur numbers have been calculated for low moduli $m\in \{1,2,3\}$. $S_1(K)=0$ trivially since for every positive integern,  $n+n+\ldots n = n \ mod \ 1$. The other two cases can be worked out and we have the following theorem \cite{modular}:

\noindent {\em Theorem 2}. The values of the modular Schur numbers for $m=2,3$ are: 
\begin{equation}
\begin{aligned}
S_2(K) &= 1, \forall K. \\
S_3(K) &=  \begin{cases} 1,  K=1,\\
2, \forall K\geq 2.
\end{cases}
\end{aligned}
\end{equation}
Therefore, the prospects of constructing quantum states satisfying modular sum-free conditions seem very positive. It is worthwhile to mention another relative of the modular Schur numbers that are also more amenable for calculations than standard Schur numbers. They are the modular weak Schur numbers. Firstly, let us introduce the notion of weak Schur numbers:
 
 \noindent {\em Definition 4}.
A set $S_w$ is said to be weakly sum-free  when  $\forall n_i, n_j \in S_w$, pairwise distinct positive integers, then 
 $n_i +  n_j = z \not\in S_w$. 
 
The following example illustrates weakly sum-free partitions for the following set of distinct integers $S=\{ 1,2,3,4,5,6,7,8 \}$ into $K=2$ partitions:
\begin{equation}\label{weakly}
S=\{ 1,2,3,4,5,6,7,8 \} = \{1,2,4,8\}_w \cup \{3,5,6,7\}_w.
\end{equation}
Notice that the subset $ \{1,2,4,8\}_w$ is weakly sum-free but not sum-free. The notion of weak Schur number read as follows:

\noindent {\em Definition 5}. Given an integer $K$, the weak Schur number denoted as $WS(K)$ is defined as the maximum integer $n$ such that the integers $1,2,\ldots ,n$ can be distributed into $K$ partitions in such a way that the $K$ partitions are weakly sum-free sets.

\noindent Since the $S(K)$ is more restricted than $WS(K)$, it follows
\begin{equation}
S(K) \leq WS(K).
\end{equation}
With exhaustive search it possible to to verify that by adding 9 to the set \eqref{weakly} does not admit any weakly sum-free partitions into $K=2$ sets.
Thus, $WS(2) = 8$. The exact values of $WS(K)$ are only known for small values of $K=1,2,3,4$ \cite{weak}:
\begin{equation}
WS(1) = 2,  WS(2) = 8,  WS(3) = 23,  WS(4) = 66.
\end{equation}

Then, the condition of being sum-free set can be decomposed into two separate contiions,
\begin{equation}
\{ S + S \not\subset S \} \equiv \{ 2S\not\subset S \} \cup \{ S + S \not\subset S \}_w
\end{equation}
where the subscript $w$ denotes a weakly sum-free condition. Notice that the selfcorrelated exclusion principle in Sec.\ref{sec:selfcorrelated} 
arises the quantization of states satisfying the condition $\{ 2S\not\subset S \}$. Likewise, the weakly sum-free condition can give rise to a correlated  exclusion principle similar to that introduced in Sec.\ref{sec:correlated}.

Then the modular weak Schur numbers are defined by combining the notions of modular Schur numbers and weak Schur numbers:

\noindent {\em Definition 6}. Given an integer $K$, the modular weak Schur number, denoted as $WS_m(K)$, is defined as the maximum integer $n$ such that the integers $1,2,\ldots ,n$ can be distributed into $K$ partitions in such a way that the $K$ partitions are weakly sum-free sets modulo $m$.

It is apparent that,
\begin{equation}
S_m(K) \leq WS_m(K).
\end{equation}
\noindent {\em Theorem 3}. The values of tthe modular Schur numbers for $m=1,2,3$ are \cite{modular}: 
\begin{equation}
\begin{aligned}
WS_1(K) &= 2K, \forall K. \\
WS_2(K) &= \begin{cases} 
2, \text{if} \ K=1, \\
4(K-1)+1, \forall K\geq 2. \\
\end{cases}\\
WS_3(K) &=  \begin{cases} 2, \text{if} \ K=1,\\
4,  \text{if} \ K=2, \\
6(K-2) + 2, \forall K\geq 3.
\end{cases}
\end{aligned}
\end{equation}
Thus, the growth of the modular weak Schur numbers is not constant in $K$ as in the modular Schur numbers but linear for small moduli, and still much less amilliorated
than the exponential growth of the original Schur numbers \eqref{exponential}. This is a good indication that quantum states could possibly be constructed with modular weakly sum-free conditions. 

\section{Conclusions}
\label{sec:conclusions}

We do not know of any phenomenon in nature supporting a principle like the
correlated exclusion principle presented here. In fact, it must be considered a mathematical axiom rather
than a physical principle, at this stage. We use the word principle only by analogy to the real exclusion principle.
But we live in the age of quantum simulations \cite{QSimulation1,QSimulation2,QSimulation3,QSimulation4,QSimulation5,QSimulation6} and the correlated exclusion principle poses a challenge in this field:
\vspace{10pt}

\noindent {\em Unsolved}: Make a proposal for an artificial physical realization of the correlated exclusion principle
and make it happen.

\vspace{10pt}

\noindent Normally, a big deal for a quantum simulation is to realize a real physical system in a regime far from the reach
of current experiments or calculations with a classical computer. The present challenge is really beyond this. It amounts to
create artificially a new quantum state of matter and a new quantum statistics. Related to this is the equivalent question: Does the correlated exclusion principle violate any fundamental law of physics? For instance, discrete symmetries like parity, charge conjugation and specially time reversal. A working rule in physics is that if something is not forbidden, it may happen.

Cold atoms in optical lattices is a very powerful platform for quantum simulations, but others could also do the job. In this regard, the quantum simulation of hard-core bosons with ultra-cold atoms \cite{Paredes} is an example of engineering in nature a type of exclusion principle that is not the originally proposed by Pauli.  Hard-core bosons are strange bosons since they are bosons when they are at different sites in space but become fermions when they are at the same site since doubly occupancy is forbidden. They are not truly fermions for the wave function does not obey the antisymmetrization principle.
The physics behind this peculiar exclusion principle is strong repulsion pair interactions between the constituents particles.
Hence, by engineering appropriately the interactions among normal particles and the lattice, we may obtain unusual quantum statistics.
The simulation of fermionic Mott insulators \cite{Esslinger} highlights an opportunity to simulate the strange fractal Mott insulator of App.\ref{sec:fractal}.
Some partial hints so as to how to realize these quantum states have been provided in Sec.\ref{sec:states} like to realize the positive integers $n_{i.k}$ by means of ordinary bosons grouped into packets of $n_{i,k}$ particles each, subject to a mechanism or interaction fulfilling the sum-free property. The nature or origin of this mechanism is unknown.

An analog quantum computation as mentioned above would have another benefit: to help compute Schur numbers from the ground state properties of the simulated new quantum states and measuring its partitions. Viceversa, any progress in the calculation of Schur numbers will benefit this type of analog quantum simulations.

The simplest instance of the selfcorrelated exclusion principle in Sec.\ref{sec:selfcorrelated} seems reachable for a quantum simulation in some optical lattice in order to engineer the fractal Mott state of App.\ref{sec:fractal}. By pursuing similar ideas, one could engineer some exotic version of a Mott insulator using superlattices and lasers that hide certain lattice sites in such a way as to rewrite the resulting system as a correlated exclusion principle of interaction-free particles.

The quasiparticles called duetons here and associated to the excitations of the correlated exclusion principle can be generalized from duets to trios, quartets etc ... by extending the congruence of two integers $n_1 + n_2 \not\in S_k$ to sum-free three congruences $n_1 + n_2 + n_3 \not\in S_k$, four congruences
$n_1 + n_2 + n_3 + n_4 \not\in S_k$, and so on and so forth. Moreover, the general ideas behind the correlated exclusion principle are far more reaching than the use of Schur numbers. In particular, similar constructions could be realized with Ramsey numbers, which are a relative of Schur numbers \cite{Weisstein}.

The possible geometrical/topological origin of the correlated exclusion principle as for fermions, bosons and anyons remains an open problem.

\begin{acknowledgments}
I thank O. Viyuela for a useful reading of the manuscript.
This work builds upon a lecture delivered at the  Real Academia de Ciencias Exactas, F\'{\i}sicas y Naturales de Espa\~{n}a \cite{RAC2016} and I thank the institution for their kind invitation to the Ciclo ``Ciencia para Todos" (2016). 
M.A.M.-D. acknowledges financial support from the Spanish MINECO, FIS 2017-91460-EXP, PGC2018-099169-B-I00 FIS-2018 and the CAM research consortium QUITEMAD+, Grant S2018-TCS-4243. The research of M.A.M.-D. has been supported in part by the U.S. Army Research Office through Grant No. W911N F- 14-1-0103.

\end{acknowledgments}

\appendix

\section{Correlated Permanents}
\label{sec:permanents}

An important difference of the quantum states constructed in Sec.\ref{sec:states}
satisfying the sum-free rule with respect to the classical partitions $S_k, 1\leq k \leq K$
is the fact that the quantum numbers are strictly ordered in increasing values whereas
the positive integers in the partitions may have arbitrary order. It is possible to construct
quantum states reinforcing the property that the order in each quantum register of the
quantum states be irrelevant by symmetrizing the states as if they were indistinguishable quantum numbers.
The way to achieve this is to introduce the permanent of a matrix and using it with the quantum registers.
The matrix we need is denoted by $A_k$ for each of the partitions $S_k$ giving rise to the quantum registers of Sec.\ref{sec:states}.
The matrix elements are constructed from the quantum numbers  $n_{i,k} $ as follows:
\begin{equation}
(A_k)_{n,i} := \ket{n_{i}}_k, \ n_{i,k} \in S_k.
\end{equation}
From this we can construct the permanent of this matrix of individual quantum states yielding a quantum register:
\begin{equation}
\ket{Perm_{\otimes}(A_k)}:= \sum_{\sigma \in S_{N_k}}\bigotimes_{i=1}^{N_k} (A_k)_{n,\sigma(i)}.
\end{equation}
Then, the complete quantum state $\ket{\Psi(K)}$ associated to the $K$ partitions of a Schur number $S(K)$ is defined as a superposition
of the quantum permanents $\ket{Perm_{\otimes}(A_k)}_{\alpha}$,
\begin{equation}
\ket{\Psi(K)}:= \frac{1}{|{\cal R}_K|}\sum_{\alpha=1}^{|{\cal R}_K|}\bigotimes_{k=1}^K\ket{Perm_{\otimes}(A_k)}_{\alpha},
\end{equation}
where $|{\cal R}_K|$ is the cardinality of the set of invariant transformations ${\cal R}_K$ of the partitions $S_k$ \eqref{set}.

\section{Partitions for $S(5)$}
\label{sec:Heule}

Exoo \cite{Exoo} provided the following first certificate showing that $S(5)\geq 160$:

\begin{widetext}
\begin{align*}
S_1&= \{1, 6, 10, 18, 21, 23, 26, 30, 34, 38, 43, 45, 50, 54, 65, 74, 87, 96, 107, 111, 116, 118, 123, 127, 131, 135, 138, 140, 143, 151, 
\\ &\quad \quad 155, 160\} \\
S_2&=\{2, 3, 8, 14, 19, 20, 24, 25, 36, 47, 51, 62, 73, 88, 99, 110, 114, 125, 136, 137, 141, 142, 147, 153, 158, 159\}\\
S_3&=\{4, 5, 15, 16, 22, 28, 29, 40, 41, 42, 48, 49, 59, 102, 112, 113, 119, 120, 121, 132, 133, 139, 145, 146, 156, 157\}\\
S_4&=\{7, 9, 11, 12, 13, 17, 27, 31, 32, 33, 35, 37, 53, 56, 57, 61, 79, 82, 100, 104, 105, 108, 124, 126, 128, 129, 130, 134, 144,
\\ &\quad \quad 148, 149, 150, 152, 154\}\\
S_5&=\{44, 52, 55, 58, 60, 63, 64, 66, 67, 68, 69, 70, 71, 72, 75, 76, 77, 78, 80, 81, 83, 84, 85, 86, 89, 90, 91, 92, 93, 94, 95, 97, 98, 
\\ &\quad \quad 101, 103, 106, 109, 117\}.
\end{align*}
\end{widetext}

Recently, M. Heule has proved that in fact $S(5)=160$ and took up 2 petabytes of space \cite{Heule}.

\noindent {\em Unsolved}: Determine the set ${\cal R}_{S(5)}$ and whether it is a group.

\section{Fractal Mott Insulator}
\label{sec:fractal}

In Sec.\ref{sec:selfcorrelated} the quantum state representing the selfcorrelated exclusion principle was
interpreted as the quantum numbers of particles with allowed values $n_i$ with
the same energy corresponding to $K=1$. As a spin-off of this result, it is appealing to interpret a similar state
but in coordinate space rather than energy representation. In this case, the quantum number $n_i$ represents
the particle located at site $n_i$ of a one-dimensional lattice. The label $i$ is now redundant,
but we shall keep it. There is an important difference between both realizations. The energy representation
satisfy a sum-free rule of type $2S\not\in S$ as represented by the existence of only one energy state, whereas
the coordinate representation does not fulfill that condition since each position state is different one another, namely
\begin{equation}
\begin{matrix}
\ket{\mathfrak{M}}_F:=\ket{1}_1\ket{3}_2 \ket{4}_3\ket{5}_4\ket{7}_5\ket{9}_6\ket{11}_7 
\ket{12}_8\ket{13}_9\\ \ket{15}_{10}\ket{16}_{11}\ket{17}_{12}\ket{19}_{13}\ket{20}_{14}\ldots \ket{\bullet}_{K=N}
\end{matrix}
\end{equation}
The peculiar feature of this localized state, a Mott state, is that it has the fractal structure in lattice space as given by the recursion formula \eqref{recursion}.


\end{document}